\documentclass{PoS}

\usepackage{amsmath}
\usepackage{subcaption}
\usepackage{units}

\title{Impact of cross section uncertainties on NOvA oscillation analyses}

\ShortTitle{Impact of cross section uncertainties on NOvA oscillation analyses}

\author{\speaker{Jeremy Wolcott}, for the NOvA Collaboration\\
        Tufts University\\
        E-mail: \email{jwolcott@fnal.gov}}

\abstract{The NOvA experiment is an off-axis long-baseline neutrino oscillation experiment measuring $\nu_{\mu}$ disappearance and $\nu_{e}$ appearance in a $\nu_{\mu}$ beam originating at Fermilab.  Oscillations are observed in the far detector relative to measurements made in the near detector, with a full simulation of the beam line and detectors being used to perform extrapolation between them.  The neutrino simulation uses the GENIE event generator, which contains implementations of numerous theoretical neutrino interaction models applied to a variety of nuclear targets.  However, recent data, recent reanalysis of extant data, and continued development of theoretical models have brought to light deficiencies in the default GENIE cross section model, which informs the predicted spectra used to infer oscillation parameters.  We explore how uncertainties in this model, together with modifications to GENIE version 2.12.2 based on external information and NOvA Near Detector (ND) data, affect NOvA's oscillation parameter inferences.  We also discuss how these uncertainties can be mitigated by judicious analysis design, including usage of the ND data.}

\FullConference{The 20th International Workshop on Neutrinos (NuFact2018)\\
		12-18 August 2018\\
		Blacksburg, Virginia}

\begin{document}

\section{Introduction}

Neutrino oscillation experiments rely heavily on predictions from Monte Carlo simulations to infer the parameters of interest from their data.  Among the most challenging components of the simulation chain for such experiments is typically a neutrino interaction generator, which predicts the rates of neutrino reactions in detector materials as well as the identities and four-momenta of reactions' outgoing particles.  The NOvA experiment, which is a long-baseline neutrino oscillation experiment based at Fermilab in Batavia, IL, currently employs the GENIE generator \cite{genie} (version 2.12.2) to simulate neutrino interactions in its near detector (ND) at Fermilab and its far detector (FD) in Ash River, MN.  

Historically, generators used by experiments, such as GENIE, have made concessions to the difficult task of predicting interactions with the complex nuclear environment by adopting a ``factorization'' approach.  In this scheme, numerous theoretical models for hard-scattering processes from hadrons or quarks at various momentum scales are composed with a relatively simple model for the nuclear dynamics.  Though this picture has been sufficient for past work, increasing statistical precision in modern experiments has begun to reveal cracks in the foundation.  In particular, dedicated measurements from neutrino scattering experiments (e.g., \cite{miniboone-qe-1, miniboone-qe-2, minerva-lowq3-1, minerva-lowq3-2, minerva-xverse-vars, t2k-incl, t2k-xverse-vars}) have cast considerable doubt on whether the non-interacting relativistic Fermi gas (RFG) nuclear model used by default in contemporary versions of GENIE is viable.  Nontrivial uncertainty is associated with both the details of the nuclear model and the hard scattering processes themselves, which rely on approximations where explicit non-perturbative calculations using QCD are untenable.

While NOvA is designed with the mitigation of these cross section uncertainties in mind, no present or planned experiment is completely insensitive to them.  In the following sections we explore the uncertainties noted above, including adjustments to GENIE's model we find we are forced to make by external data, improvements to available theory, and our own ND data.  We then discuss the impact they have on NOvA's $\nu_{\mu}$ disappearance and $\nu_{e}$ appearance measurements, given NOvA's design---the detectors are built to be as similar as possible in materials and technologies---and the calorimetric energy reconstruction principle used in the analyses.

\section{GENIE 2.12.2 model and adjustments}
\begin{figure}[htb]
	\centering
	\begin{subfigure}{0.45\textwidth}
		\includegraphics[width=\textwidth,trim={0 0.25cm 0 0.25cm},clip]{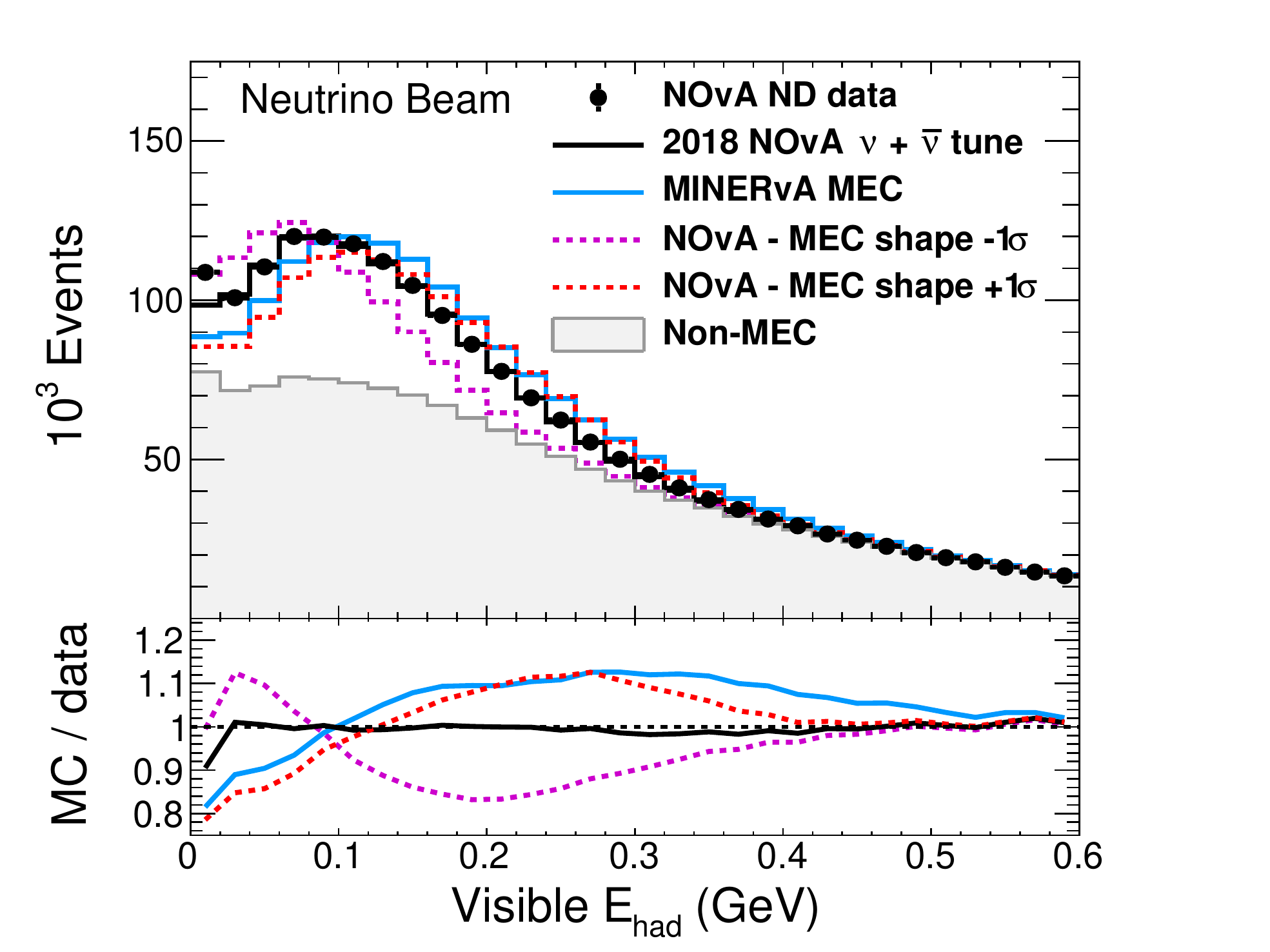}
	\end{subfigure}
	\begin{subfigure}{0.45\textwidth}
		\includegraphics[width=\textwidth,trim={0 0.25cm 0 0.25cm},clip]{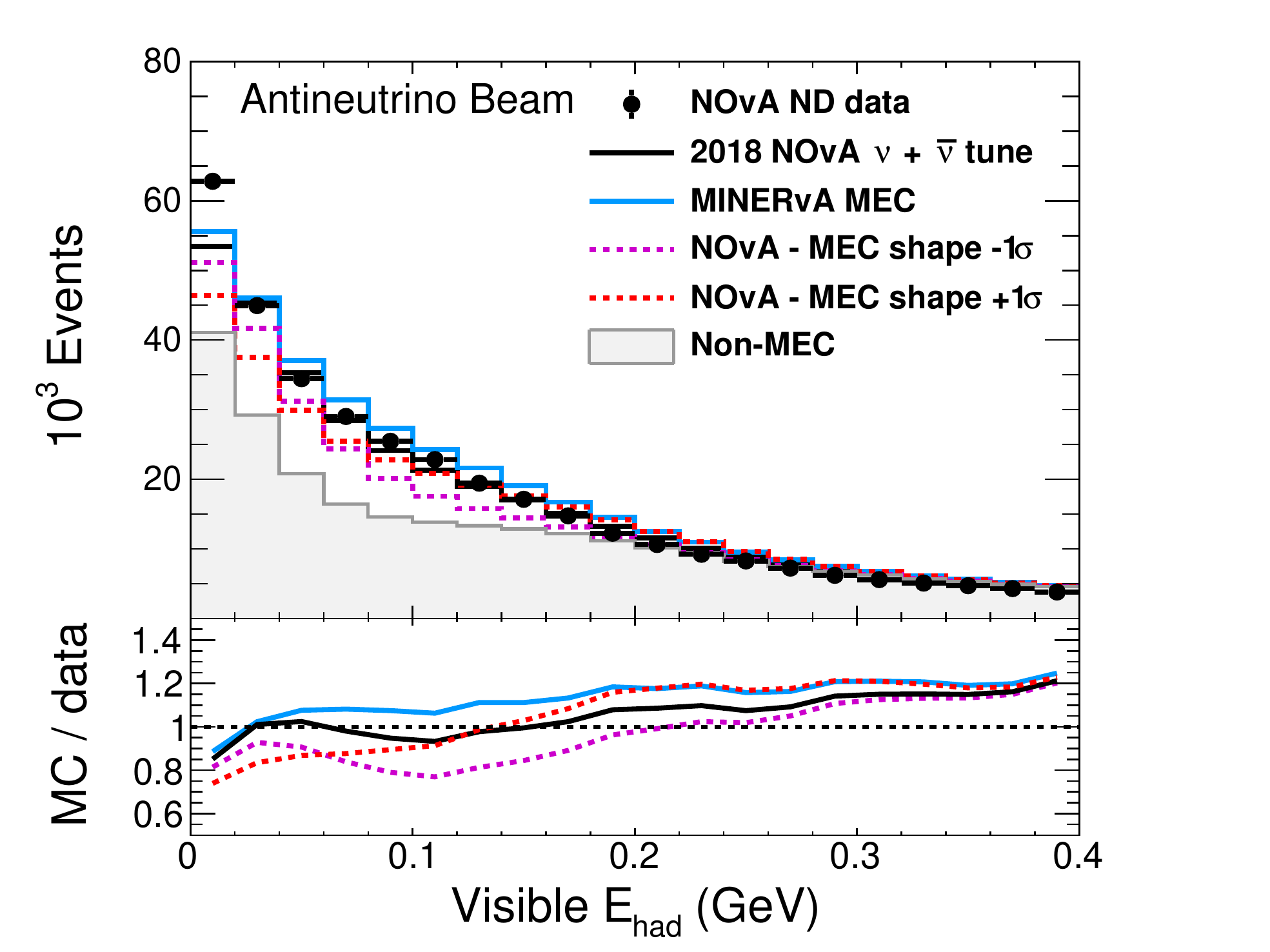}
	\end{subfigure}
	\caption{Distributions in visible hadronic energy (event energy deposited in the scintillator less that deposited by the muon) for $\nu_{\mu}$ CC candidates in the NOvA ND.  Black points are data; the shaded histogram is non-MEC prediction, while colored curves represent the gray component summed with various predictions for MEC: solid black is our central value; dotted curves represent our uncertainties, arising from fits to our data with different assumptions about the non-MEC prediction; solid blue uses MINERvA's tuned MEC prediction referenced in the text.}
	\label{fig:MEC tuning}
\end{figure}

GENIE 2.12.2's default model divides the total hard-scattering cross section into numerous processes for which independent models exist.  The largest ones (and the only ones we will consider here) in charged-current (CC) interactions are, in order of increasing final-state hadronic mass $W$, quasielastic scattering (QE), resonant baryon production (RES), and nonresonant deep inelastic scattering (DIS).  In current analyses, NOvA makes adjustments to the axial mass in the QE dipole form factor (setting $M_A = \unit[1.04]{GeV}$ rather than the default $\unit[0.99]{GeV}$) and nonresonant single pion production with $W < \unit[2.0]{GeV}$ (reducing it to 43\% of its nominal value) based on reanalysis of bubble chamber data that these parameters were originally tuned to \cite{bubble-qe, bubble-nonres-reana}.  These make relatively small differences in the prediction.  A much larger impact comes from the addition of a new hard-scattering process, that of two-nucleon ejection via a meson-exchange current (MEC) process.\cite{nustec-whitepaper}  Because this is a reaction well known from electron scattering, but no contemporary model is able to describe the extant neutrino data\cite{minerva-lowq3-1, minerva-lowq3-2, minerva-xverse-vars, t2k-xverse-vars}, NOvA has elected to enable the optional ``Empirical MEC'' model in GENIE\cite{katori-empMEC} and tune it to NOvA ND data in energy- and three-momentum transfer $(q_0, |\vec{q}|)$.  Comparisons of the default and tuned predictions to NOvA ND data, as well as the uncertainties constructed from alternative tunes, and the outcome of a similar procedure performed by the MINERvA Collaboration to their own data\cite{minerva-mec-tune}, are shown in fig. \ref{fig:MEC tuning}.

As noted above, the most precarious component of the current generator prediction is the nuclear model.  NOvA alters the default GENIE 2.12.2 model in several ways to address shortcomings here.  First, there is widespread agreement that long-range interactions of the nuclear potential between nucleons affect QE reactions, significantly suppressing them at low $Q^2$ and mildly enhancing them at higher $Q^2$ relative to the RFG prediction.  We adopt the random phase approximation (RPA)-based calculation of the Val\`{e}ncia group \cite{valencia-rpa} parameterized as reweights in $(q_0, |\vec{q}|)$ to the GENIE QE model by R. Gran \cite{gran-rpa} and the associated uncertainties.  Measurements of delta resonant production in external data \cite{miniboone-res, minos-qe, minerva-pi-1, minerva-pi-2}, as well as our own ND data, also suggest the presence of nuclear dynamics resulting in a similar suppression at low $Q^2$ relative to the free nucleon prediction as the RPA effect, so we also apply the $Q^2$ parameterization of the RPA effect to RES as a placeholder for whatever the true nuclear effect may be.  We take the unmodified RES prediction as an uncertainty variation.

\section{Impact on neutrino oscillation measurements}

As discussed elsewhere in these Proceedings \cite{erica-nufact}, NOvA uses a calorimetric technique for both $\nu_{\mu}$ and $\nu_e$ energy reconstruction in which neutrino energy is estimated using a function of both lepton and hadronic system energies.  Uncertainties in cross section modeling can impact the fidelity of these estimators in a number of different ways: for instance, shifting the balance of energy between the better-resolved leptonic and the more-poorly-resolved hadronic systems in CC events; changing the predicted mean energy that is unseen by the detector (due either to the assumed nuclear binding potential or hadronic energy that escapes as neutrons) and must be added back by the estimator; or adjusting the expected frequency of background processes that have different energy responses than the signal.

To mitigate the impact of these uncertainties on the prediction at the FD, NOvA relies on measurements at the ND, which are propagated to predictions for the FD via an ``extrapolation'' procedure.  The latter supposes that discrepancies observed between ND simulation and data distributions can be accounted for in the FD prediction by modifying the ND true event rate in bins of true energy, which can then be multiplied by the simulated ratio of the geometric and oscillation effects between the two detectors to yield the FD true rate.  This is conveniently expressed as a matrix equation over the energy bins:
\begin{equation}
	\vec{N}_{FD} = \vec{N}_{ND}\ \mathbf{R}\ \mathbf{M}_{ND}\ \mathbf{F}\ \mathbf{P}_{osc}\ \mathbf{M}_{FD}^{-1}
	\label{eq:fovern}
\end{equation}
Here, the $\vec{N}_{\alpha}$ are the predicted event yields in bins of reconstructed energy for detector $\alpha$; the diagonal matrix $\mathbf{R}$ contains the bin-by-bin ratios of the observed and predicted ND yields, $R_{ii} = N^{ND}_{\mathrm{data},i} /N^{ND}_{\mathrm{MC},i}$; the $\mathbf{M}_{\alpha}$ are so-called ``migration'' matrices between reconstructed and true energies for detector $\alpha$, from simulation; the diagonal $\mathbf{F}$ is denoted the ``far over near ratio,'' $F/N$, which encodes the predicted effect of the neutrino beam dispersion and the difference in acceptance between the detectors; and the diagonal $\mathbf{P}_{osc}$ applies oscillation probabilities for given oscillation parameters.  This approach differs from the strategy sometimes employed by other oscillation experiments in which parameters in the model are fitted to the ND data and propagated to the FD prediction via their fitted covariance matrix.  While the NOvA strategy is less general (it is only effective when the ND and FD share the same underlying cross section uncertainties, like in NOvA, for instance), it is guaranteed to reproduce the observed ND distribution, even if unknown effects are present in the data that the model cannot account for.

\begin{figure}[htb]
	\centering
	\begin{subfigure}{0.45\textwidth}
		\includegraphics[width=\textwidth,trim={0 0 0 0.3cm},clip]{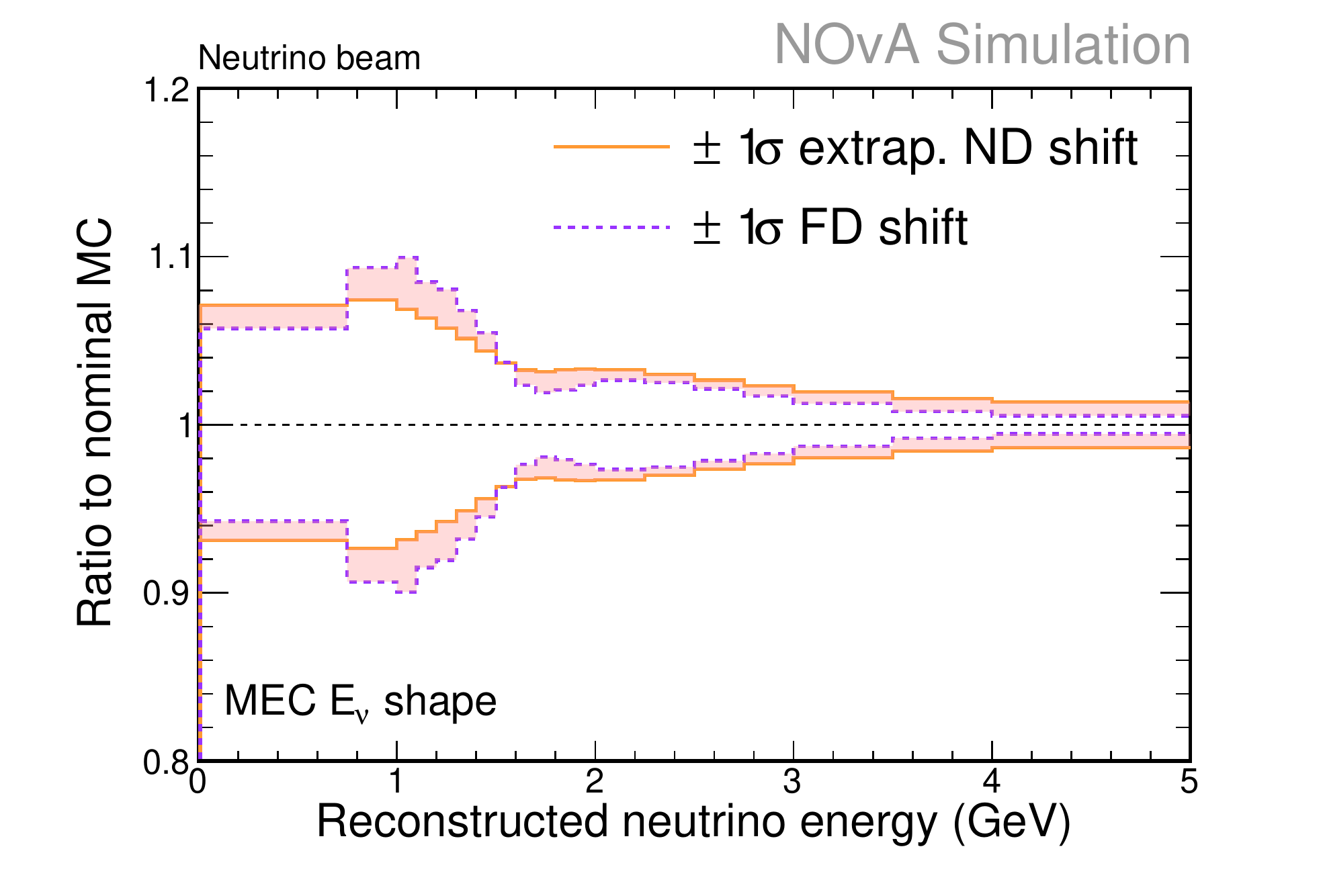}
	\end{subfigure}
	\begin{subfigure}{0.45\textwidth}
		\includegraphics[width=\textwidth,trim={0 0 0 0.3cm},clip]{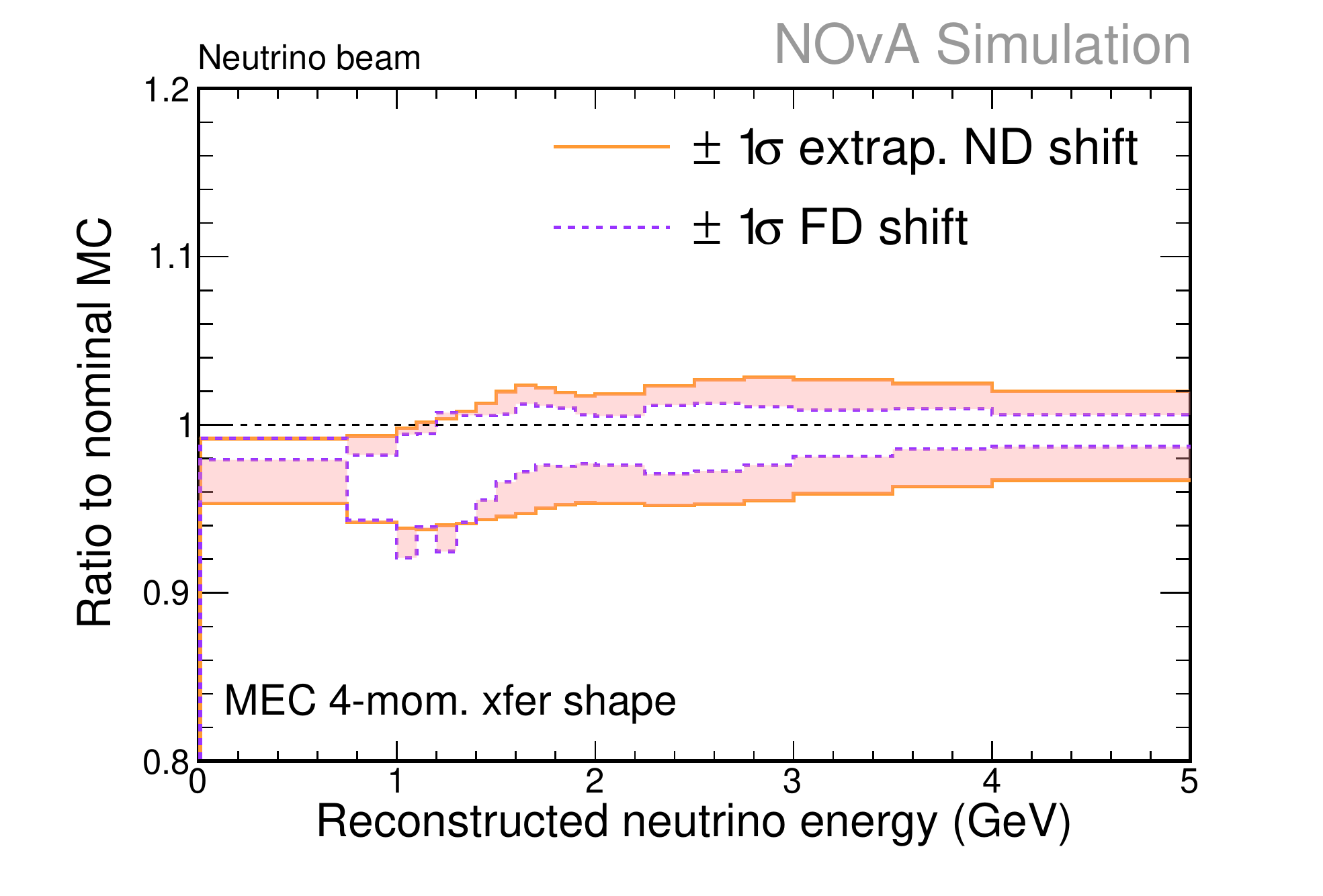}
	\end{subfigure}
	\caption{Ratio of predicted FD CC $\nu_{\mu}$ yields $\vec{N}_{FD}'$ (orange) and $\vec{N}_{FD}''$ (purple) for $\pm 1\sigma$ shifts to the nominal in two important uncertainties in the MEC model: its shape as a function of $E_{\nu}$, left, and its shape in $(q_{0},|\vec{q}|)$, right.  (See text for the definitions of the $\vec{N}$.)  The shaded residual difference between the two predictions corresponds to the uncertainty \textit{not} accounted for by extrapolation.}
	\label{fig:extrap MEC unc}
\end{figure}

The extent to which the $F/N$ method enables calculation of the effect of changes in the cross section model on the FD prediction using ND data can be illustrated with test cases.  In such a test, the ND data is replaced by a modified prediction using a designated cross-section change during the calculation of $\mathbf{R}$, resulting in a modified $\mathbf{R}'$.  The $\vec{N}_{FD}'$ obtained from applying eq. \ref{eq:fovern} to $\mathbf{R}'$ can then be compared to a different $\vec{N}_{FD}''$ obtained by directly applying the modified cross section model to the FD prediction in simulation.  If $\vec{N}_{FD}'$ and $\vec{N}_{FD}''$ coincide, then the extrapolation procedure can perfectly account for the effect of the given cross-section shift using the ND data.  If they differ, on the other hand, the residual between $\vec{N}_{FD}''$ (direct FD prediction under shifted model) and $\vec{N}_{FD}'$ (extrapolation of shifted prediction with nominal model) illustrates the fraction of the given shift that is not ``canceled'' (i.e., is left uncorrelated between the two predictions) by the extrapolation procedure.  Fig. \ref{fig:extrap MEC unc} shows the comparison resulting from shifts due to two important uncertainties in the MEC model noted above; the extrapolation procedure reduces the original uncertainties of up to 10\% to a few percentage points.

\begin{figure}[htb]
	\centering
	\begin{subfigure}{0.45\textwidth}
		\includegraphics[width=\textwidth,trim={0 0 0 0.3cm},clip]{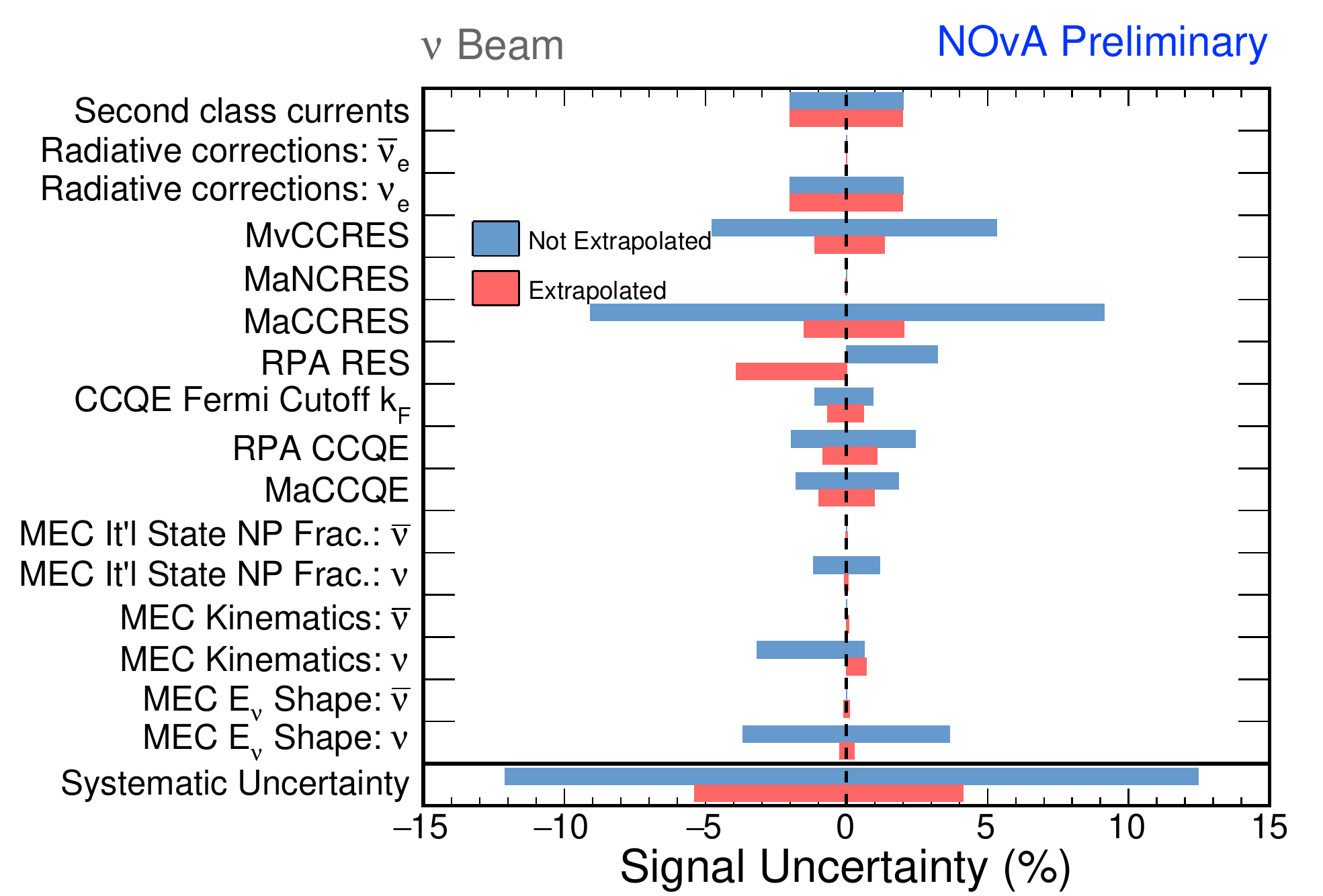}
	\end{subfigure}
	\begin{subfigure}{0.45\textwidth}
		\includegraphics[width=\textwidth,trim={0 0 0 0.3cm},clip]{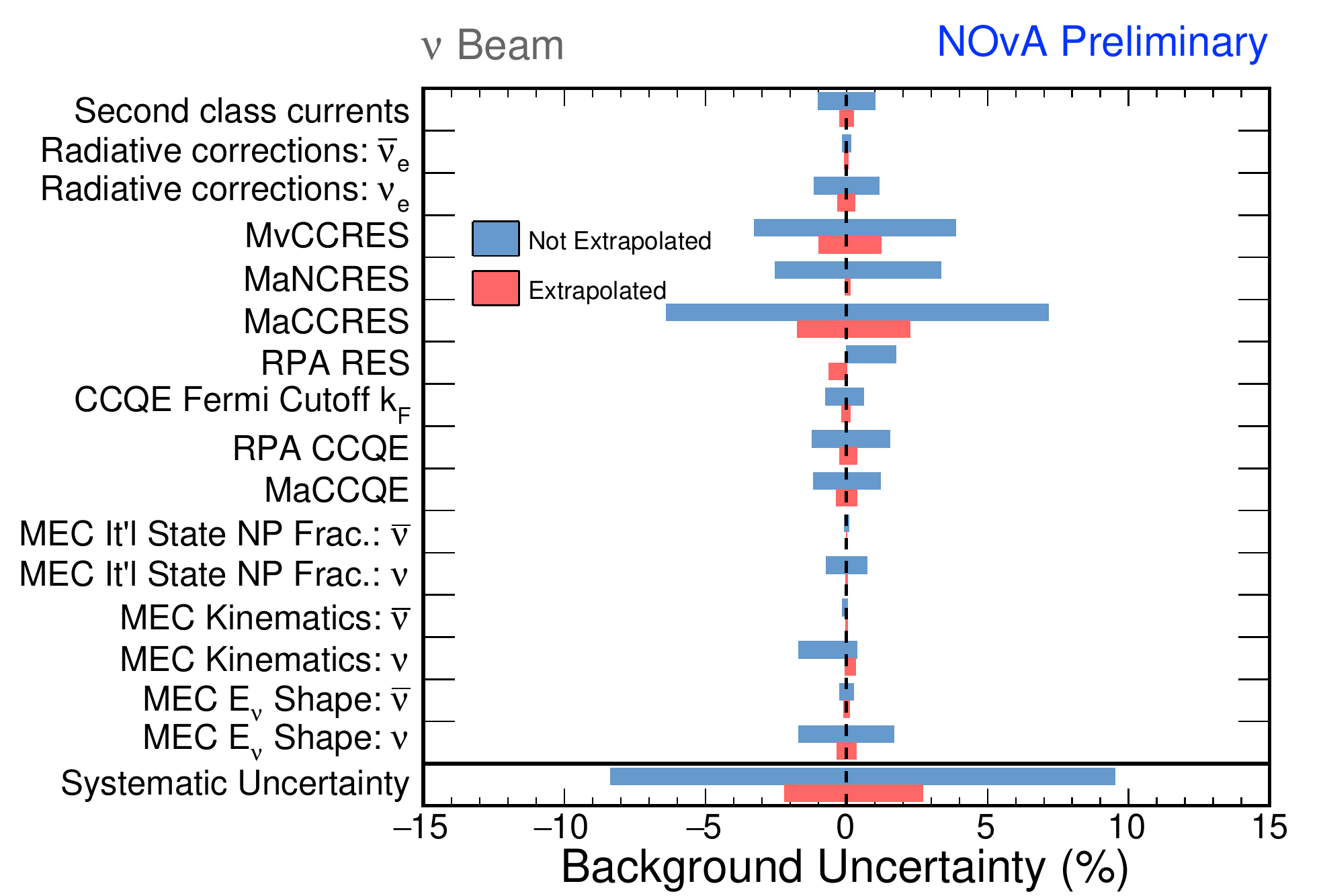}
	\end{subfigure}
	\caption{Effect of selected cross section uncertainties on the signal (left) and background (right) predictions for the $\nu_e$ appearance measurement, before (blue) and after (red) extrapolation.}
	\label{fig:extrap xsec nue}
\end{figure}

In fits to the FD data, the extrapolation procedure is used first to correct the nominal FD prediction.  Known uncertainties are then accounted for using nuisance parameters constructed from bin-by-bin splines fitted to the difference between shifted predictions $\vec{N}_{FD}''$ and the corrected nominal prediction.  The reduction of the cross section impact on uncertainties in the $\nu_{e}$ signal and background predictions due to extrapolation is illustrated in fig. \ref{fig:extrap xsec nue}.  Even after extrapolation is applied, however, neutrino cross section uncertainties retain significant influence on the results, together accounting for 35\%, 44\%, and 53\% of the total systematic error budgets for NOvA's $\sin^2(\theta_{23})$, $\Delta m_{32}^2$, and $\delta_{CP}$ measurements, respectively.  We anticipate that future continued improvements to cross section modeling, particularly in regard to the nuclear dynamics in QE and RES interactions, the detailed nature of 2p2h, and antineutrino reactions, will be essential as the statistical precision of these measurements improves and systematics begin to limit them.

\section{Conclusions}
NOvA relies on strong internal constraints on cross section uncertainties for its oscillation program derived from the functionally identical detector paradigm and a calorimetric neutrino energy reconstruction technique.  In addition, a comprehensive program is underway to ensure that all relevant cross section issues are considered.  After the constraint from the ND is applied, cross section uncertainties currently comprise 30-50\% of the systematic budget on the most important oscillation parameter measurements.  We look forward to continued development of models and associated systematic treatments in the community, new measurements of cross sections to help constrain them, and ultimately their integration into improved oscillation parameter measurements in NOvA.

\end{document}